\documentclass[twocolumn,aps,pre,fleqn,showpacs]{revtex4}
\usepackage{amsmath,amssymb,graphicx}
\usepackage{times,mathptm}

\newcommand{\mean}[1]{\langle{#1}\rangle}
\newcommand{\ud}{\mathrm{d}}
\newcommand{\dt}{\mathrm{d}t}
\newcommand{\kc}{k_{\mathrm{c}}}
\newcommand{\ktc}{\tilde{k}_{\mathrm{c}}}
\newcommand{\km}{k_{\mathrm{min}}}
\newcommand{\mk}{\mean{k}}
\newcommand{\kM}{k_{\mathrm{max}}}
\newcommand{\dynAA}{A + A \rightarrow \emptyset}
\newcommand{\dynAB}{A + B \rightarrow \emptyset}
\newcommand{\qaak}[1][]{Q^{\mathrm{(aa)#1}}_k}
\newcommand{\qbbk}[1][]{Q^{\mathrm{(bb)#1}}_k}
\newcommand{\qabk}[1][]{Q^{\mathrm{(ab)#1}}_k}
\newcommand{\qaa}[1][]{Q^{\mathrm{(aa)#1}}}
\newcommand{\qbb}[1][]{Q^{\mathrm{(bb)#1}}}
\newcommand{\qab}[1][]{Q^{\mathrm{(ab)#1}}}
\newcommand{\qba}[1][]{Q^{\mathrm{(ba)#1}}}

\newcommand{\rhok}{\rho_k}
\newcommand{\rhoa}{\rho^{\mathrm{(a)}}}
\newcommand{\rhob}{\rho^{\mathrm{(b)}}}
\newcommand{\rhoc}{\rho_{\mathrm{c}}}
\newcommand{\na}{n^{\mathrm{(a)}}}
\newcommand{\nb}{n^{\mathrm{(b)}}}
\newcommand{\tha}{\Theta^{\mathrm{(a)}}}
\newcommand{\thb}{\Theta^{\mathrm{(b)}}}
\newcommand{\Vs}{\mathcal{V}}

\begin{document}

\title{Multicomponent reaction-diffusion processes on complex networks}

\author{Sebastian~Weber}
\affiliation{Institut~f\"ur~Festk\"orperphysik,
             Technische~Universit\"at~Darmstadt,
             Hochschulstr.~8, 64289~Darmstadt, Germany}

\author{Markus~Porto}
\affiliation{Institut~f\"ur~Festk\"orperphysik,
             Technische~Universit\"at~Darmstadt,
             Hochschulstr.~8, 64289~Darmstadt, Germany}

\date{May 9, 2006}

\begin{abstract}
  We study the reaction-diffusion process $\dynAB$ on uncorrelated scale-free
  networks analytically. By a mean-field ansatz we derive analytical expressions
  for the particle pair-correlations and the particle density. Expressing the
  time evolution of the particle density in terms of the instantaneous particle
  pair-correlations, we determine analytically the `jamming' effect which arises
  in the case of multicomponent, pair-wise reactions. Comparing the relevant
  terms within the differential equation for the particle density, we find that
  the `jamming' effect diminishes in the long-time, low-density limit. This even
  holds true for the hubs of the network, despite that the hubs dynamically
  attract the particles.
\end{abstract}

\pacs{89.75.Hc, 82.20.-w, 05.40.-a}

\maketitle

\section{Introduction}
\label{sec:introduction}

The merging of graph theory and methods from classical statistical physics has
led to the modern theory of complex networks
\citep{albert-statMechNet02,dorogotvtsev-evolNets03}. This development went
along with the discovery of complex networks in a vast variety of sciences.
Almost all observed real networks share a so called scale-free degree
distribution. That is, the number of links (the degree $k$) of a randomly chosen
vertex of the network is distributed according to
\begin{equation}
  \label{eq:scaleFreeDef}
  P(k) \propto k^{-\gamma} .
\end{equation}
For exponents $2 < \gamma < 3$ this distribution has a diverging second moment
$\mean{k^2}$ for infinite size networks ($N \rightarrow \infty$). A very
prominent example is the WWW \citep{satorras-inet04} with a value $\gamma$ of
approximately $2.1$. As a consequence, the network is very robust against the
removal of random nodes, but extremely weak upon removal of the highly connected
nodes \citep{cohen-inetBreak00}.

The diverging $\mean{k^2}$ leads to enormous fluctuations in the degree of a
vertex and causes a large heterogeneity in the vertex connectivity. This crucial
feature of scale-free networks has strong consequences on dynamical processes
taking place on them. The absence of an epidemic threshold in the presence of a
diverging $\mean{k^2}$ is one of the most notable examples
\citep{boguna-epid03,boguna-mechanics03}.

A similar class of dynamical processes, the reaction-diffusion processes
\citep{torney-dyn83,toussaint-dyn83}, has been studied numerically on networks
by \citeauthor{gallos-kin04} \citep{gallos-cor05,gallos-kin04} and shortly after
also analytically within mean-field (MF) theory by
\citeauthor{catanzaro-annAA05} \citep{catanzaro-annAA05} (see
\citep{hastings-seriesExp06} for a recent series-expansion approach to
networks). This type of dynamics is capable of modeling epidemic spreading,
chemical reactions, and many more. The analytical work by
\citeauthor{catanzaro-annAA05} calculated the density decay of the one component
$\dynAA$ reaction for homogeneous and heterogeneous complex networks. The
results for homogeneous networks agree with the classical MF behavior of a
density decay linear in time, $1/\rhoa(t) \propto t$. In the case of
heterogeneous (i.e., scale-free) networks, one finds a more general power law
for the density decay
\begin{equation}
  \label{eq:rhoMeanAA}
  \frac{1}{\rhoa(t)} \propto t^{\alpha(\gamma)} \left( \ln t
  \right)^{\beta(\gamma)} ,
\end{equation}
with the $\gamma$-dependent characteristic exponents given by
\begin{align}
  \label{eq:rhoMeanAAexpon1}
  \alpha(\gamma) &=
  \begin{cases}
    1/(\gamma - 2) & \quad 2 < \gamma < 3 \\
    1 & \quad \gamma \geq 3
  \end{cases}
\end{align}
and
\begin{align}
  \label{eq:rhoMeanAAexpon2}
    \beta(\gamma) &=
  \begin{cases}
    1 & \quad \gamma = 3 \\
    0 & \quad \text{otherwise}
  \end{cases} .
\end{align}
The reason for this extremely fast density decay originates in the existence of
a few vertices with a very large degree (so-called hubs) in a scale-free
network. As the analysis shows, the density is constantly $1/2$ for vertices
with a degree $k > \kc = \mk / 2\rhoa(t)$.

If one wants to extend these results to more complicated diffusion-annihilation
dynamics, for instance the $\dynAB$ process, which up to now has been studied
only numerically \citep{gallos-kin04,gallos-cor05}, one realizes that this
diffusion-annihilation dynamics introduces forbidden diffusion steps, since a
vertex can hold at most one particle at a time. This constraint prohibits an $A$
($B$) particle to diffuse to an adjacent vertex which is already occupied by
another $A$ ($B$) particle. As we are going to discuss below, this leads to a
`jamming' effect, a new feature of the dynamical process. Nevertheless, the
obtained solution has a structure similar to that of the $\dynAA$ process. The
paper begins in Section \ref{sec:dynAB} with a MF calculation of the particle
pair-correlations in dependence of the particle density. It follows a MF ansatz
for the density decay of the $\dynAB$ process from which we quantify the
`jamming' effect. The results of the analytical work are compared with numerical
simulations in Section \ref{sec:numerics}. At last, we conclude with Section
\ref{sec:conclusions}.

\section{The $\dynAB$ reaction on complex networks}
\label{sec:dynAB}

We assume the complex network of $N$ nodes to be fully defined by its $N \times
N$ adjacency matrix $a_{ij}$. To discuss a physically meaningful complex network
in the sense of diffusion, we take the network to be undirected and free of
self- and multiple-connections. Therefore, $a_{ij}$ is a traceless, symmetric
binary matrix with the elements $a_{ij}$ being $0$ or $1$, which symbols a
(dis)connection between site $i$ and $j$. The state of vertex $i$ at time $t$ is
described by two dichotomous variables $\na_i(t)$ and $\nb_i(t)$. Their values
may be $1$ or $0$ only, indicating the presence or absence of a particle $A$
$(\na_i(t))$ and $B$ $(\nb_i(t))$. The system state is thus defined by
\begin{equation}
  \label{eq:sysDefAB}
  \begin{split}
  \mathbf{n}(t) &= \mathbf{\na}(t) + \mathbf{\nb}(t) \\
    \mathbf{\na}(t) &= \{\na_1(t), \na_2(t), ..., \na_N(t)\}\\
    \mathbf{\nb}(t) &= \{\nb_1(t), \nb_2(t), ..., \nb_N(t)\}.
  \end{split}
\end{equation}
Note, that these variables have to fulfill the constraint $\na_i(t) \, \nb_i(t)
= 0$ at any time. In the course of the calculation, we will take the average
over multiple realizations of the same system, turning the discrete $\na_i(t)$
and $\nb_i(t)$ variables into densities $\rhoa_i(t)$ and $\rhob_i(t)$.
Furthermore, we will assume throughout the analysis the statistical equivalence
of vertices of the same degree $k$. Therefore, denoting by $\Vs(k)$ the set of
all vertices with the same degree $k$, we assume that
\begin{equation}
  \label{eq:statEquivAB}
  \begin{split}
  \rhoa_i(t) \equiv \rhoa_k(t) &\quad \forall \, i \in \Vs(k) \\
  \rhob_i(t) \equiv \rhob_k(t) &\quad \forall \, i \in \Vs(k) \\
  \end{split}
\end{equation}
is valid. Following standard MF treatment, we hence neglect all fluctuations
which might exist within a set of vertices $\Vs(k)$. The total density
$\rhoa(t)$ ($\rhob(t)$) is given by the set of partial particle densities
$\{\rhoa_k(t)\}$ ($\{\rhob_k(t)\}$) through the relation
\begin{equation}
  \label{eq:totDensDef}
  \rhoa(t) = \sum_k \rhoa_k(t) \, P(k) .
\end{equation}
A dynamics starts by random assignment of maximal one particle per vertex. The
particles diffuse by random jumps at a rate $\lambda$ to adjacent neighbors
through the network. If two different particles meet at a vertex, they instantly
annihilate and the vertex becomes empty. Before we proceed to the time evolution
of the system, we first derive an expression for the particle pair-correlations
in dependence of the partial particle densities.

\subsection{Particle pair-correlations}
\label{sec:corr}

We quantify the particle pair-correlations for given partial particle densities
by counting the number of contacts between particles on adjacent vertices. To
count the $AB$ contacts of a vertex $i$, we assume vertex $i$ to carry an $A$
particle and count the number of adjacent vertices which are occupied by a $B$
particle. Setting this number in relation to all connections of the vertex $i$
yields the pair-correlation coefficient
\begin{equation}
  \label{eq:corrAnsatz}
  q^{\mathrm{(ab)}}_i(t) = \frac{1}{k_i} \na_i(t) \sum_j a_{ij} \nb_j .
\end{equation}
Averaging now over a whole ensemble of equal systems and making use of the usual
MF assumption $\mean{\na_i(t) \, \nb_j(t)} \approx \mean{\na_i(t)}
\mean{\nb_j(t)}$, we obtain
\begin{equation}
  \label{eq:corrAveraged}
  Q^{\mathrm{(ab)}}_i(t) = \frac{1}{k_i} \rhoa_i(t) \sum_j a_{ij} \rhob_j(t) .
\end{equation}
By using the statistical equivalence of all $N_k$ vertices $i$ with the same
degree $k$, we can sum over all these vertices such that
\begin{equation}
  \label{eq:corrCalcul}
  \qabk(t) = \frac{1}{k} \rhoa_k \sum_{k'} \rhob_{k'}(t)
  \frac{1}{N_k} \sum_{i \in \Vs(k)} \sum_{j \in \Vs(k')}
    a_{ij} .
\end{equation}
In this step, we split the sum with index $j$ into two sums over $k'$ and one
over $\Vs(k')$. The double sum over $a_{ij}$ is related to the conditional
probability $P(k'|k)$ that a vertex of given degree $k$ has a neighbor which has
degree $k'$. This equation has been derived previously
\citep{boguna-mechanics03} to be
\begin{equation}
  \label{eq:matrixProbRel}
  \frac{1}{k N_k} \sum_{i \in \Vs(k)} \sum_{j \in \Vs(k')} a_{ij} = P(k'|k) .
\end{equation}
Using this equation and assuming an uncorrelated network, which simplifies the
conditional probability $P(k'|k)$ to $k' P(k') / \mk$, we obtain the expression
\begin{equation}
  \label{eq:corrQabk}
  \qab_k(t) = \rhoa_k(t) \, \thb(t) ,
\end{equation}
where we define
\begin{equation}
  \label{eq:thetaDef}
  \thb(t) = \frac{1}{\mk} \sum_{k'} k' \rhob_{k'}(t) \, P(k') .
\end{equation}
One should note that by introducing the mean $\mean{k}$, the values of the
exponent $\gamma$ are limited to $\gamma > 2$. Otherwise the mean $\mean{k}$ is
not defined in the limit of infinite size networks ($N \to \infty$). The overall
particle pair-correlation coefficient $\qab(t)$ can easily be computed by
multiplying Eq.~\eqref{eq:corrQabk} with $P(k)$ and summing once more over all
$k$,
\begin{equation}
  \label{eq:corrTotal}
  \qab(t) = \rhoa(t) \, \thb(t) .
\end{equation}
Analogously, we have $\qaa(t) = \rhoa(t) \, \tha(t)$, $\qbb(t) = \rhob(t) \,
\thb(t)$, and $\qba(t) = \rhob(t) \, \tha(t)$.

\subsection{Density decay}
\label{sec:densDecay}

For further computations we will assume for simplicity that the initial
densities of $\rhoa$ and $\rhob$ are equal, so that there is a symmetry between
$A$ and $B$ particles. We will calculate only an expression for $\na(t)$, and
one may obtain the corresponding $\nb(t)$ equations by interchanging indices $A$
and $B$. Modeling the diffusion as a Poisson process \citep{kampen-stoch92}, the
set of $\{\na_i(t)\}$ changes within an infinitesimal time interval $\dt$ as
\begin{widetext}
\begin{equation}
  \label{eq:ansatzAB}
  \na_i(t + \dt) = \na_i(t) \, \eta^{\mathrm{(a)}}_i(\dt) 
  + \left[ 1 - (\na_i(t) +
    \nb_i(t)) \right] \xi^{\mathrm{(a)} }_i(\dt) .
\end{equation}
Here $\eta^{\mathrm{(a)} }_i$ and $\xi^{\mathrm{(a)} }_i$ are dichotomous random
variables, taking values of $0$ or $1$ with certain probabilities $p$ and $1-p$
respectively,
\begin{align}
  \label{eq:etaDefAB}
  \eta^{\mathrm{(a)}}_i(\dt) & =
   \begin{cases}
    0 & \; \displaystyle p = \lambda \, \dt \left[ \sum_j \frac{a_{ij}
        \nb_j(t)}{k_j} + \left( 1 - \frac{1}{k_i} \sum_j
        a_{ij} \na_j(t) \right) \right] \\
    1 & \; 1 - p
  \end{cases} \\
 \label{eq:xiDefAB}
  \xi^{\mathrm{(a)} }_i(\dt) & =
  \begin{cases}
    1 & \; \displaystyle p = \lambda \, \dt \sum_j \frac{a_{ij}
      \na_j(t)}{k_j} \\
    0 & \; 1-p
  \end{cases} .
\end{align}
The following two cases need to be distinguished: (i)~If site $i$ is occupied by
an $A$ particle at instant $t$, $\eta^{\mathrm{(a)} }_i(\dt)$ is responsible for
the next time step: The site may become empty ($\eta^{\mathrm{(a)} }_i = 0$)
with a probability proportional to the product of the jumping rate $\lambda$ and
the time interval $\dt$ if a $B$ particle in the neighborhood jumps onto site
$i$ or if the $A$ particle at $i$ jumps away to a neighborhood site where no $A$
particle is already located. Otherwise no change happens. (ii)~If the site $i$
is empty at instant $t$, then $\xi^{\mathrm{(a)} }_i(\dt)$ will determine the
time evolution: The vertex may become occupied by an $A$ particle only if one in
the neighborhood jumps onto vertex $i$. Note that the two random variables
$\eta^{\mathrm{(a)}}_i$ and $\xi^{\mathrm{(a)}}_i$ are hence not independent
from each other, but we will treat them as independent (cf.\
\citep{catanzaro-annAA05}).

Equation~\eqref{eq:ansatzAB} yields an average time evolution for $\na_i(t)$
\begin{equation}
  \label{eq:naMeanAB}
  \begin{split}
    \mean{\na_i(t + \dt)} & = \na_i(t) - \dt \Biggl\{ \na_i(t) \\
    & \quad + \sum_j
    \biggl[ \na_i(t) \frac{a_{ij} \nb_j(t)}{k_j} - \na_i(t)
    \frac{1}{k_i} a_{ij} \na_j(t) 
    - \left( 1 - \left[ \na_i(t) + \nb_i(t) \right] \right)
    \frac{a_{ij}\na_j(t)}{k_j} \biggr] \Biggr\} ,
  \end{split}
\end{equation}
where we have set without loss of generality the jumping rate $\lambda = 1$.
Averaging over a whole set of equal initial configurations and applying once
more Eq.~\eqref{eq:matrixProbRel} and the statistical equivalence of vertices
with the same degree, Eq.~\eqref{eq:statEquivAB}, we obtain after some formal
rearrangements
\begin{equation}
  \label{eq:rhoKMeanFAB}
    \frac{\ud \rhoa_k}{\dt} = - \rhoa_k 
    - \sum_{k'} \Biggl\{ \frac{1}{k'} \left[
      \rhoa_k \rhob_{k'} - k' \rhoa_k \frac{1}{k} \rhoa_{k'} - \rhoa_{k'} +
      \rhoa_k \rhoa_{k'} + \rhob_k \rhoa_{k'} \right]
    k P(k'|k)\Biggr\} .
\end{equation}
Here we have suppressed the explicit time dependence for the sake of simplicity.
Assuming the network to be uncorrelated (i.e.\ that $P(k'|k) = k' \, P(k')/\mk$)
allows us to perform the sum over $k'$, yielding finally the expression
\begin{equation}
  \label{eq:rhoKMeanSAB}
    \frac{\ud \rhoa_k}{\dt} = - \rhoa_k 
    - \frac{k}{\mean{k}} \left[
      \rhoa_k \rhob - \rhoa + \rhoa_k \rhoa + \rhob_k \rhoa
    \right] 
    + \rhoa_k \tha
\end{equation}
\end{widetext}
for the partial particle densities. Multiplying Eq.~\eqref{eq:rhoKMeanSAB} with
$P(k)$ and summing over all $k$ values results in the differential equation for
the overall density,
\begin{equation}
  \label{eq:rhoMeanA_AB}
  \begin{split}
  \frac{\ud \rhoa}{\dt} & = - \rhob \tha - \rhoa \thb \\
  & = - \qab - \qba .
  \end{split}
\end{equation}
From Eq.~\eqref{eq:rhoMeanA_AB} it is apparent that the density decay is
directly proportional to the pair-correlations among unlike particles. To
proceed further, we need expressions for $\rhoa_k$ and $\rhob_k$. Since the
initial densities are equal, we have forcibly $\qaa = \qbb$ because of symmetry.
This implies the equality $\rhoa_k = \rhob_k \equiv \rho_k'$, allowing further
simplifications and transforming Eq.~\eqref{eq:rhoKMeanSAB} into
\begin{equation}
  \label{eq:rhoKEqAB}
  \begin{split}
    \frac{\ud \rho_k'}{\dt} & = - \rho_k' + \frac{k}{\mean{k}} \left[
      1 - 3 \rho_k'
    \right] \rho'
    + \rho_k' \Theta' \\
    & = - \rho_k' + \frac{k}{\mean{k}} \left[
      1 - 3 \rho_k'
    \right] \rho'
    + Q_k'' .
    \end{split}
\end{equation}
This differential equation is very similar to the one previously found for the
$\dynAA$ process \citep{catanzaro-annAA05}, with an additional term
$Q_k''~\equiv~\qaak~=~\qbbk$ and a coefficient of $3$ instead of $2$ in front of
$\rho_k'$. The additional term measures the number of contacts among particles
of the same type, which slow down the reaction as these jumps are prohibited,
causing a `jamming' effect.

To test whether this new term $Q_k'' = \rho_k' \Theta'$ alters the behavior of
the dynamics, we compare it to the other term $3 k \rho_k' \rho'/\mean{k}$,
which is as well quadratic in the density, and find
\begin{equation}
  \label{eq:estimate}
  \frac{\Theta'}{3 \rho' k/\mean{k}} =
  \frac{1}{3 k} \frac{\sum_{k'} k' \rho_{k'}' P(k')}{
    \sum_{k'} \rho_{k'}' P(k')} =
  \frac{1}{3 k} \frac{\mean{k' \rho_{k'}'}}{\mean{\rho_{k'}'}} .
\end{equation}
Due to the fact that the particles are dynamically attracted by the hubs of the
network, which has been shown analytically in Ref. \citep{catanzaro-annAA05} for
the $\dynAA$ process and numerically for the $\dynAB$ process in Ref. 
\citep{gallos-kin04} as well, we propose the following approximation to proceed:
We know from the $\dynAA$ process that hubs ($k > \kc$) drive the dynamics, and
that the density on those hubs is almost constant, $\rhoc = 1/2$, whereas the
density on vertices which are not hubs is substantially lower. If we approximate
the densities of all vertices with a degree $k < \kc$ to be zero, the terms
$\mean{k' \rho_{k'}'}$ and $\mean{\rho_{k'}'}$ become in the thermodynamic limit
\begin{align}
  \label{eq:estimateCalcul1}
  \mean{k \rho_{k}'} & \approx \rhoc' \sum_{k = \kc}^\infty k P(k)
  \approx \rhoc' \int_{\kc}^\infty k^{1-\gamma} \, \ud k = \rhoc'
  \frac{\kc^{2-\gamma}}{\gamma-2} \\
  \label{eq:estimateCalcul2}
  \mean{\rho_k'} & \approx \rhoc' \sum_{k = \kc}^\infty P(k) \approx
  \rhoc' \int_{\kc}^\infty k^{-\gamma} \, \ud k = \rhoc'
  \frac{\kc^{1-\gamma}}{\gamma-1} .
\end{align}
In this step, we apply the continuous $k$ approximation, which allows us to
replace the sum by an integral. By inserting
expression~\eqref{eq:estimateCalcul1} and \eqref{eq:estimateCalcul2} into
Eq.~\eqref{eq:estimate}, we find
\begin{equation}
  \label{eq:estimateResult}
  \frac{\Theta'}{3 \rho' k/\mean{k}}
  \approx \frac{1}{3} \frac{\gamma-1}{\gamma-2} \frac{\kc}{k} .
\end{equation}
Therefore, we can neglect $Q_k''$ for a vertex with $k \gg \kc$. For $k \ll
\kc$, we can neglect $Q_k''$ (and $3 k \rho_k' \rho'/\mk$) as this term is
quadratic in the density, being very small for nodes with $k \ll \kc$ in the
long-time, low-density limit. The intermediate range of $k \approx \kc$ is
difficult to assess analytically and needs to get quantified in the next Section
by numerical simulations, which show that expression \eqref{eq:estimate} is
substantially smaller than $1$ in the low density limit even for the range where
$k \approx \kc$. Therefore, `jamming' is only of relevance for vertices with a
low degree and high densities.

The calculation leading to Eq.~\eqref{eq:estimateResult} is carried out in the
limit of infinite size network. In all real networks, one inevitably has a
maximum degree $\kM$ inducing finite-size effects. This maximum degree $\kM$
limits the upper bound of the integrals in Eqs.~\eqref{eq:estimateCalcul1} and
\eqref{eq:estimateCalcul2}. Evaluating these integrals with such an upper bound,
one obtains
\begin{equation}
  \label{eq:estimateFiniteResult}
  \frac{\Theta'}{3 \rho' k/\mean{k}} \approx
  \frac{1}{3} \frac{\gamma-1}{\gamma-2} \frac{\kc}{k} \, f(\kc/\kM),
\end{equation}
with
\begin{equation}
  \label{eq:scalingFunc}
  f(x) = \frac{1 - x^{\gamma-2}}{1 -
    x^{\gamma-1}} .
\end{equation}
The scaling function $f(x)$ has the limiting values
\begin{equation}
  \label{eq:scalingLimits}
  f(x) =
  \begin{cases}
    1 & x \rightarrow 0 \\
    \displaystyle \frac{\gamma - 2}{\gamma - 1} & x \rightarrow 1 .
  \end{cases}
\end{equation}
Since $f(x)$ is a monotonically decreasing function for $\gamma > 2$, the
finite-size effect on the result in Eq.~\eqref{eq:estimateResult} is to slightly
decrease the importance of the `jamming' term evenly for all degrees $k$.

Concluding that the `jamming' effect is not relevant in the long-time,
low-density limit, we neglect the $Q_k'' = \rho_k' \Theta'$ term in
Eq.~\eqref{eq:rhoKEqAB} and obtain an equation which relates $\rho_k'$ to its
derivative. However, we aim at a relation for $\rho_k'$ itself. As done for the
$\dynAA$ process \citep{catanzaro-annAA05}, we proceed with the quasi-static
approximation, setting $\ud \rho_k' / \dt \approx 0$. This assumes that the
diffusion process is at any time much faster than the annihilation reaction (see
also \citep{Noh-randomWalk06}). This approximation should be valid in the case
of low densities, when the complex network is sparsely populated and the number
of diffusion events in a time interval $\dt$ is much larger than the
annihilation events, implying that the particles are always in an equilibrium
state with respect to the degree distribution. Doing so, we get an approximate
expression for $\rho_k'$,
\begin{equation}
  \label{eq:rhoKABapp}
  \rho_k' = \frac{\rho' k/\mean{k}}{1 + 3 \rho' k/\mean{k}} .
\end{equation}
This expression for $\rho_k'$ has the same structure as found for the $\dynAA$
process except that the coefficient of $\rho'$ in the denominator is $3$ instead
of $2$. As the structure of the differential equations is the same as for the
$\dynAA$ process, we obtain the same scaling-behavior of
Eqs.~\eqref{eq:rhoMeanAAexpon1},\eqref{eq:rhoMeanAAexpon2} for each component. 
The new critical $\kc$ for which a vertex is sensed as a hub by the dynamics is
$\kc = \mean{k}/ 3 \rho' = 2 \mean{k}/ 3 \rho$. Therefore, for the hubs in the
system with $k > \kc$, Eq.~\eqref{eq:rhoKABapp} is close to $1/3$, which is
completely consistent with MF, as this means that hubs are occupied by
approximately $1/3$ of the time by each component $A$ and $B$, and are empty for
the remaining $1/3$ of the time.

\section{Numerical simulations}
\label{sec:numerics}

To test the analytical results obtained, we performed intensive numerical
simulations of the $\dynAB$ process on scale-free networks. The uncorrelated,
scale-free networks are generated with the uncorrelated configuration model
(UCM) algorithm \citep{molloyReed-lcrg,molloyReed-crit,catanzaro-ucm05} and have
a size of $N = 10^6$ if not stated otherwise. The exponents $\gamma$ simulated
are in the range of $2.1$ to $3.5$, while we only present a suitable subset in
the figures. In short, one draws for a network of size $N$ a random number for
each vertex according to the degree distribution $P(k) \propto k^{-\gamma}$,
with an upper cut-off $\kM = N^{1/2}$ to ensure that the generated network is
uncorrelated \citep{boguna-cutoff04}. The drawn number corresponds to the target
degree of each node and can be understood as half-edges to be joint with other
half-edges to form a connection. This is done in the central loop, in which one
draws randomly two half-edges and joins them if this neither creates a
self-connection nor a multiple-connection. Upon a successful join of two
half-edges, these two half-edges are dropped from the set of eligible
half-edges. In any case, one continues with the central loop by drawing again
two half-edges, and so forth. After the distribution of all half-edges, we only
keep the largest component of the generated network. We use a minimum degree of
$\km = 2$ in our simulations, so that the largest component usually coincides
with the full network. We have verified that the networks are indeed
uncorrelated by obtaining the degree-degree correlation coefficient
\citep{newman-mix02}, which has an absolute value smaller than $10^{-3}$ in all
cases. On these networks, the dynamics is simulated in the following way:
Initially a fraction $\rho = 2 \rhoa = 2 \rhob$ of randomly chosen vertices is
selected, which we choose as $\rho = 0.1$ or $0.95$. Then, the algorithm assigns
randomly an equal amount of $A$ and $B$ particles to the set of chosen vertices.
After this initial setup, the diffusion-annihilation dynamics starts. First, a
vertex which carries a particle and a random adjacent neighbor of this vertex
are randomly selected. Three cases need to be distinguished: (i)~If the neighbor
vertex is empty, the particle moves to the new vertex, leaving the initial
vertex empty. (ii)~If the neighbor vertex is occupied by a particle of the other
type, an annihilation reaction occurs and both vertices become empty. 
Accordingly, the number of particles is decreased for each particle type by one,
$\na \rightarrow \na - 1$, and $\nb \rightarrow \nb - 1$. (iii)~If the neighbor
vertex is occupied by a particle of the same type, then no jump occurs. In any
case, the time is updated by $t \rightarrow t + 1/(\na + \nb)$, where $\na$ and
$\nb$ correspond to the values before the diffusion step, and one continues by
selecting randomly another vertex carrying a particle, and so forth.

In order to obtain the system's typical behavior, we average over $50$
independent dynamics on each graph and over $100$ independent graphs, making up
a total of $5000$ dynamics per data-point.

\subsection{Validation of the approximations}
\label{sec:validation-approx}

\begin{figure}[t]
  \includegraphics[scale=0.8]{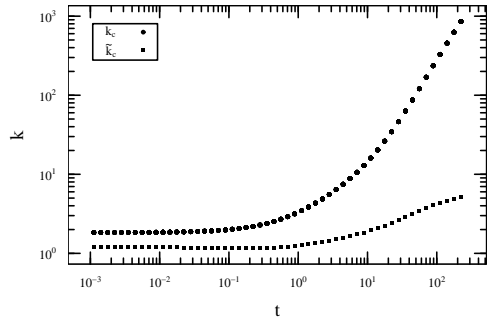}
  \caption{Plot of $\kc$ and $\ktc$ of the numerically simulated $\dynAB$
    process for an exponent $\gamma = 2.75$ and an initial density $\rho_0 =
    0.95$. Since $\kc$ is increasing much faster than $\ktc$, `jamming' becomes
    quickly irrelevant.}
  \label{fig:jam-ktc}
\end{figure}

\begin{figure}[t]
  \includegraphics[scale=0.8]{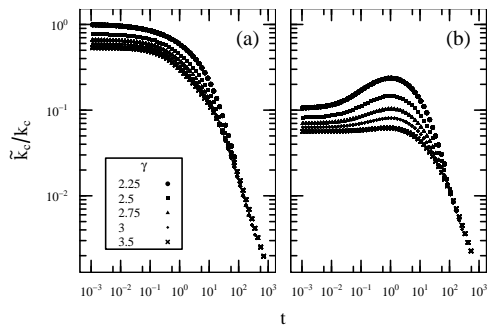}
  \caption{Ratio of $\ktc$ and $\kc$ of the numerically simulated $\dynAB$
    process on networks of different exponents $\gamma$ as indicated, with
    (a)~an initial particle density $\rho_0 = 0.95$ and (b)~$\rho_0 = 0.1$.}
  \label{fig:jam-approx}
\end{figure}

To validate the analytical calculations developed in the last section, we have
to verify the two central approximations made which are based on the assumption
of a small particle density on the network. Furthermore, it is crucial to get an
estimate which densities can be considered small enough for the validity of the
approximations. Our first approximation was to neglect in
Eq.~\eqref{eq:rhoKMeanSAB} the `jamming' term $Q_k''$ in comparison to the other
term quadratic in the density $3 k \rho_k' \rho'/\mean{k}$. We have shown the
validity of this approximation analytically for vertices with a degree $k \gg
\kc$ and $k \ll \kc$ (the latter for low densities). To check the intermediate
range $k \approx \kc$, we perform numerical simulations. If we set the ratio in
Eq.~\eqref{eq:estimate} equal to $1$, we obtain a critical degree $\ktc$,
\begin{equation}
  \label{eq:jamCritK}
  \ktc = \frac{\Theta'}{3 \rho' / \mean{k}},
\end{equation}
which separates vertices whose `jamming' term is less important than the other
quadratic density term in Eq.~\eqref{eq:rhoKEqAB} from those vertices for which
the `jamming' term is at least of equal importance. The time-evolution of the
particle density on vertices with a degree $k \gg \ktc$ is not affected by
`jamming', whereas vertices with a degree of the order of $\ktc$ or lower are
affected. On the other hand vertices with a small degree do not contribute to
the overall particle density at later times, since the hubs dynamically attract
the particles and carry the highest density $\rho_c' = 1/3$. Considering only
vertices as hubs which have a degree $k > \kc = \mean{k}/3\rho'$, we have as a
condition for `jamming' not being relevant
\begin{equation}
  \label{eq:jamCond}
  \ktc \ll \kc .
\end{equation}
If condition \eqref{eq:jamCond} is fulfilled, there are no vertices left in the
network which do carry a sufficiently high density and whose `jamming' term is
important for the time-evolution of their $\rho_k(t)$. In Fig. \ref{fig:jam-ktc}
we exemplified this condition for an exponent $\gamma = 2.75$ and an initial
particle density $\rho_0 = 0.95$. Note that the curves are only drawn until
$\kc$ reaches the value of the maximum degree $\kM$ present in the network. Once
the value of $\kc$ is much greater than $\ktc$ the `jamming' effect is of no
more relevance for the time evolution of the process for all vertices in the
network, including those with $k \approx \kc$. It is crucial to note that $\kc$
grows much faster than $\ktc$ in the course of the process. Therefore, the
`jamming' is continuously diminishing during the dynamics and only important for
low degree vertices carrying a high density in the beginning of the process. An
equivalent criterion to test weather `jamming' is not relevant is to check if
the ratio of $\ktc / \kc$ is substantially smaller than $1$. In
Fig.~\ref{fig:jam-approx}(a) we illustrate this for an initial density $\rho_0 =
0.95$. Again, the individual curves are only drawn until $\kc$ reaches $\kM$.
They start with a maximum value of almost $1$, indicating the presence of
`jamming' and drop quite quickly well below $1$. In Fig.~\ref{fig:jam-approx}(b)
we show the same simulations starting but with a much smaller initial density of
$\rho_0 = 0.1$. Most importantly, these curves already begin at values well
below $1$ and therefore there is never `jamming' present in the dynamics. The
interesting intermediate increase of $\ktc/\kc$ for the initial density $\rho_0
= 0.1$ (Fig.~\ref{fig:jam-approx}(b)) comes from the fact the dynamical hubs
start with a density $\rho_0' = 0.05$ which is smaller than their long-time
density $\rhoc' = 1/3$. Therefore, all vertices with $\rho_k' < 1/3$ and a
degree $k > \kc$ will have increasing particle densities $\rho_k'$ which enter
$\Theta'$ in Eq.~\eqref{eq:jamCritK}. Once the dynamics has reached its
long-time behavior, there are no more $\rho_k'$-terms in $\Theta'$ which
increase in magnitude, since the dynamical hubs carry the highest density in the
network.

\begin{figure}[t]
  \includegraphics[scale=0.8]{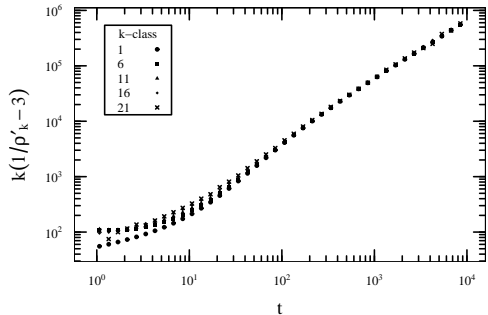}
  \caption{Numerical validation of the quasi-static approximation according to
    Eq.~\eqref{eq:static-approx} for the $\dynAB$ process, exemplified by an
    exponent $\gamma = 2.5$ and an initial density $\rho_0 = 0.1$. The
    $k$-classes are logarithmically joined, where class $1$ corresponds to
    vertices of degree $k = 2$, class $6$ to $8 \leq k \leq 10$, class $11$ to
    $33 \leq k \leq 42$, class $16$ to $134 \leq k \leq 178$, and class $21$ to
    $563 \leq k \leq 750$. A data-collapse is observed for $t \gtrsim 50$.}
  \label{fig:static-approx}
\end{figure}

The second approximation made to obtain an expression for $\rhok'$, the
quasi-static assumption $\ud \rhoa_k / \dt \approx 0$, yielded
Eq.~\eqref{eq:rhoKABapp}. Rearranging Eq.~\eqref{eq:rhoKABapp} into
\begin{equation}
  \label{eq:static-approx}
  k \, \left( \frac{1}{\rhok'} - 3 \right) = \frac{\mk}{\rho'}
\end{equation}
leads to an expression where the right hand side (and consequently the left hand
side as well) is independent of $k$ if the approximation is indeed valid. 
Plotting the left hand side of Eq.~\eqref{eq:static-approx} for a couple
different degrees $k$ should yield a data-collapse onto a single curve.
Fig.~\ref{fig:static-approx} illustrates this in the case of an exponent $\gamma
= 2.5$. The curves join quite nicely at roughly $t \approx 50$. Similar time
points are obtained for other exponents $\gamma$. We can therefore expect that
the quasi-static approximation holds after this time.

\subsection{Density decay and pair-correlations}
\label{sec:densitiy-dec-and-corr}

The verification of the scaling behavior for each component as predicted by
Eq.~\eqref{eq:rhoMeanAAexpon1} and Eq.~\eqref{eq:rhoMeanAAexpon2} turns out to
be a hard numerical problem, as finite-size effects occur quite early. A
detailed discussion and derivation of finite-size effects for the $\dynAA$
process can be found in Ref.~\citep{catanzaro-annAA05}. The typical `scale-free'
behavior of the dynamics corresponds to a density decay as a power-law of the
time with an exponent larger than $1$. \citeauthor{catanzaro-annAA05}
\citep{catanzaro-annAA05} showed that the characteristic `scale-free' behavior
of the dynamics is driven by the `dynamical' hubs of the system, where
`dynamical' hubs are those vertices which have a degree $k > \kc \propto
1/\rhoa$. As the density is decreasing monotonically in time, the number of
`dynamical' hubs decreases as well. Thus, if in a finite network of size $N$
there are no `dynamical' hubs left, the density decay turns over to a decay
linear in time. The vertex with the largest degree $\kM$ therefore limits the
duration until the cross-over from `scale-free' to non-`scale-free' behavior
happens. On the other hand, the largest degree $\kM$ being possible for
uncorrelated scale-free networks which contain neither self- nor
multiple-connections scales with the square root of the system size, $\kM
\propto N^{1/2}$, making very large system sizes necessary.

\begin{figure}[t]
  \includegraphics[scale=0.8]{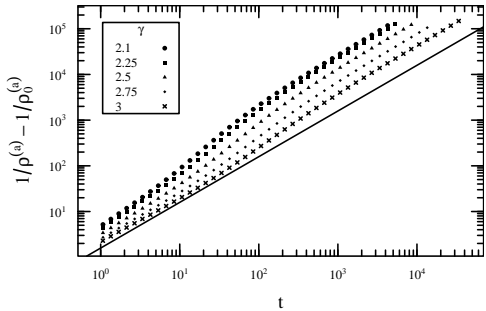}
  \caption{Density decay of the numerically simulated $\dynAB$ process for an
    initial density $\rho_0 = 0.1$ and different exponents $\gamma$. With
    increasing exponent $\gamma$, the density decay behaves `scale-free' typical
    for a longer period of time. The solid line has a slope of $1$ and is shown
    as a guide to the eye.}
  \label{fig:densDecrAB2}
\end{figure}

\begin{figure}[t]
  \includegraphics[scale=0.8]{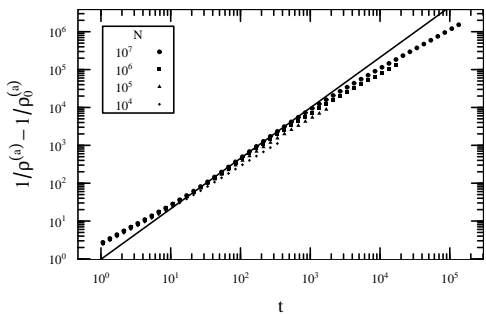}
  \caption{Density decay of the numerically simulated $\dynAB$ process for
    networks of various sizes, exemplified by an exponent $\gamma = 2.75$ and an
    initial density $\rho_0 = 0.1$. The plot illustrates the strong finite-size
    effects of the dynamics. For $\gamma = 2.75$ the process is sufficiently
    slow to show the theoretic slope of $4/3$ (drawn as a solid line) for large
    networks with size $N = 10^7$ for two decades.}
  \label{fig:densDecrAB}
\end{figure}

One might expect that the scaling exponent derived in MF theory for each
component's density decay, $\alpha(\gamma) = 1/(\gamma - 2)$, is missing some
effects which slow down the reaction for exponents $\gamma$ close to $2$. 
Otherwise, the diverging $\alpha(\gamma)$ for $\gamma \rightarrow 2$ would
result in a diverging reaction speed. Forcibly, we expect to recover the scaling
law given by $\alpha(\gamma)$ for $\gamma \rightarrow 3$. Choosing a value of
$\gamma$ which is smaller than but close to $3$ has the convenient side effect
that the density decay is relatively slow, such that the dynamics will show a
`scale-free' behavior for a longer period of time with the appropriate density
decay exponent larger than $1$. We verify this assumption by simulating the
$\dynAB$ process for various exponents $\gamma$ while keeping the system size
constant at $N = 10^6$. In Fig.~\ref{fig:densDecrAB2} we show the resulting
density decays. The curves deviate from a linear in time decay only for very
short durations, but the amount of time with which each process behaves
`scale-free' increases with increasing exponent $\gamma$. To recover the scaling
behavior of $\rhoa$, we chose an exponent $\gamma = 2.75$ which corresponds to a
value $\alpha = 4/3$. In Fig.~\ref{fig:densDecrAB}, we present the results for
the density decay $\rhoa$ for various network sizes of up to $N = 10^7$.
Nevertheless, even for a system size of $N = 10^7$, the density decay shows
`scale-free' behavior only for less than two decades.

\begin{figure}[t]
  \includegraphics[scale=0.8]{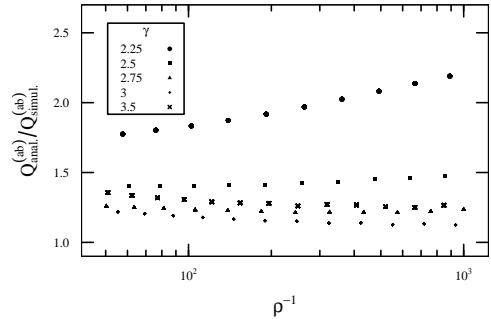}
  \caption{Ratio between the analytically calculated correlations
    $\qab_{\mathrm{anal.}}$ and the numerically obtained one
    $\qab_{\mathrm{simul.}}$, plotted as a function of the inverse density
    $\rho^{-1}$ for an initial density $\rho_0 = 0.1$ and different exponents
    $\gamma$. The range of the inverse density shown is limited by the validity
    of the quasi-static approximation, which is valid at sufficiently low
    densities, $\rho^{-1} > 5 \times 10^1$. However, the inverse density is
    also limited from above to $\rho^{-1} < 10^3$, as the pair-correlation is
    quadratic in the particle density. Such observables can only be
    meaningfully resolved as long as $\rho > N^{-1/2}$, which yields the
    condition $\rho > 10^{-3}$ for the given system size of $N = 10^6$.}
  \label{fig:correl-ana-ab}
\end{figure}

A far more sensitive alternative check of the analytical calculations is given
by the direct comparison of the analytically calculated particle
pair-correlations, Eq.~\eqref{eq:corrTotal}, and the measured ones from
numerical simulations, as these particle pair-correlations directly drive the
annihilation rate. Therefore, we evaluate the ratio between the analytical
predictions of Eq.~\eqref{eq:corrTotal} and the corresponding numerically
obtained values for the pair-correlations. It is crucial to note that $\qab(t)$
can be understood as a function of all partial particle densities $\rhoa_k(t)$
and $\rhob_k(t)$, so that it depends only implicitly via these densities on time
$t$. Approximating these partial particle densities via the quasi-static
assumption, one may write the particle pair-correlations as a function of the
inverse particle density $\rho^{-1}$ by using Eq.~\eqref{eq:rhoKABapp},
\begin{equation}
  \label{eq:corrTotalApprox}
  \qab_{\text{anal.}}(\rho^{-1}) = \frac{1}{2 \rho^{-1}} \, \frac{1}{\mk} \,
  \sum_{k'} k' \,
  \frac{k'/\mk}{2 \rho^{-1} + 3 k'/\mk} \, P(k') .
\end{equation}
The inverse particle density $\rho^{-1}$ can be regarded as an alternative
measure of time, since $\rho^{-1}$ is a monotonically increasing function in the
course of the dynamics. In Fig.~\ref{fig:correl-ana-ab}, we present the
resulting ratios for various exponents $\gamma$. The expected value of $1$ for
the ratio, which would mean perfect agreement of analytical and numerical
particle pair-correlations, is quite well achieved for exponents $\gamma$ close
to $3$. Smaller exponents $\gamma$ yield ratios somewhat larger than $1$, which
however depend only weakly on $\rho^{-1}$. The reason for the larger ratios is
presumably that these networks have larger fluctuations in the network
connectivity structure. These topological fluctuations may lead to strong
density fluctuations which are not captured by the current MF ansatz. For
$\gamma > 3$, the ratios become somewhat larger than $1$ as well, as it is seen
for $\gamma = 3.5$. This behavior results from a segregation of the components,
which is well known for lattices \citep{toussaint-dyn83} and has already been
observed by \citeauthor{gallos-kin04} \citep{gallos-kin04} for networks with
$\gamma > 3$. That is, for networks with $\gamma > 3$ the dynamics behaves as on
lattices, which is as well reflected by the scaling relations from
Eq.~\eqref{eq:rhoMeanAAexpon1} and Eq.~\eqref{eq:rhoMeanAAexpon2} giving a
linear in time density decay for networks with $\gamma > 3$. Such segregation is
not captured in our MF calculation of the pair-correlations and is therefore out
of range of its validity.

\section{Conclusions}
\label{sec:conclusions}

This paper presents a detailed discussion of the $\dynAB$ process on networks.
We show analytically the existence of a `jamming' effect for this two component
reaction and quantify it analytically with the correlations among unlike
particles. The analysis can easily be generalized to multicomponent, pairwise
processes and be used to derive a MF theory for a process where $n$ particle
types react pairwise with each other. Note, however, that the applicability is
limited to cases where it is guaranteed that the particle densities are of the
same order for all times. Otherwise, one component of the system with the
largest density might cause a non-negligible `jamming' and could significantly
slow down the reaction. For the $\dynAB$ process discussed here, we show that
`jamming' is only important for vertices with small degree $k$ and high
densities. In the long-time, low-density limit we derive analytically that
`jamming' vanishes for all vertices in the network, including the hubs. This
conclusion is supported by numerical simulations and allows us to reason that
the particle densities of the $\dynAB$ process on scale-free, uncorrelated
network show for each component the same scaling behavior as the one of the
$\dynAA$ process.

\bibliography{ab_mft}

\end{document}